%% file: ms.tex
\documentclass[chicago]{emulateapj_jw}

\usepackage{hyperref}
\usepackage{epsfig}
\usepackage{natbib}
\usepackage{graphicx}                
\usepackage{lscape}
\usepackage{verbatim}              
\usepackage{amsmath,amsthm,amsfonts,amsopn,amssymb} 
\usepackage{longtable}
\usepackage{pdflscape}
\usepackage{afterpage}
\usepackage{rotating}					
\usepackage{ulem}
\usepackage{epstopdf}  
\usepackage{multirow}
\usepackage{color}
\usepackage[left]{lineno}
\newcommand{\totaltargets}{138 }

\newcommand{\detectstar}{42 } 
\newcommand{\detectsys}{35 } 
\newcommand{\rvstar}{22 }

\newcommand{\myaopalomar}{68 }
\newcommand{\myaokeck}{5 }
\newcommand{\myaototal}{73 }
\newcommand{\myaototalno}{65 }

\received{}
\accepted{}

\slugcomment{to appear in ApJ}
\shorttitle{}
\shortauthors{Wang}

\begin{document}
\title{Influence of Stellar Multiplicity On Planet Formation. IV. Adaptive Optics Imaging of \emph{Kepler} Stars With Multiple Transiting Planet Candidates}
\author{
Ji Wang\altaffilmark{1},
Debra A. Fischer\altaffilmark{1},
Ji-Wei Xie\altaffilmark{2},
David R. Ciardi\altaffilmark{3},
} 
\email{ji.wang@yale.edu}
\altaffiltext{1}{Department of Astronomy, Yale University, New Haven, CT 06511 USA}
\altaffiltext{2}{Department of Astronomy \& Key Laboratory of Modern Astronomy and Astrophysics in Ministry of Education, Nanjing University,
210093, China}
\altaffiltext{3}{NASA Exoplanet Science Institute, Caltech, MS 100-22, 770 South Wilson Avenue, Pasadena, CA 91125, USA}

\begin{abstract}


The \emph{Kepler} mission provides a wealth of multiple transiting planet systems (MTPS). The formation and evolution of multi-planet systems are likely to be influenced by companion stars given the abundance of multi stellar systems. We study the influence of stellar companions by measuring the stellar multiplicity rate of MTPS. {{We select \totaltargets bright (K$_P< 13.5$) \emph{Kepler} MTPS and search for stellar companions with AO imaging data and archival radial velocity (RV) data. We obtain new AO images for \myaototal MTPS. Other MTPS in the sample have archival AO imaging data from the \emph{Kepler} Community Follow-up Observation Program (CFOP). }}From these imaging data, we detect \detectstar stellar companions around \detectsys host stars. For stellar separation $1\ \rm{AU}<a<100$ AU, the stellar multiplicity rate is $5.2\pm5.0$\% for MTPS, which is 2.8$\sigma$ lower than $21.1\pm2.8$\% for the control sample, i.e., the field stars in the solar neighborhood. We identify two origins for the deficit of stellar companions within 100 AU to MTPS: (1) a suppressive planet formation, and (2) the disruption of orbital coplanarity due to stellar companions. To distinguish between the two origins, we compare the stellar multiplicity rates of MTPS and single transiting planet systems (STPS). However, current data are not sufficient for this purpose. For $100\ \rm{AU}<a<2000$ AU, the stellar multiplicity rates are comparable for MTPS ($8.0\pm4.0$\%), STPS ($6.4\pm5.8$\%), and the control sample ($12.5\pm2.8$\%). 

\end{abstract}


\section{Introduction}
\label{sec:intro}

As exoplanet surveys reach higher sensitivity and longer time baseline, more exoplanets are being discovered. Many of these exoplanets are in multi-planet systems. {{As of September 2015, the radial velocity (RV) technique and the transit method have detected 152 and 857 planets in multi-planet systems~\citep[http://exoplanets.org,][]{Han2014}}}. {{From these systems, we can study their orbital spacing~\citep[e.g.,][]{Wright2011,Burke2014}, mutual inclination~\citep[e.g.,][]{Lissauer2011,Tremaine2012}, and eccentricity distribution~\citep[e.g.,][]{Juric2008, Kane2012, Xie2015}.}} These studies can be used to test theories of planet formation and dynamical evolution~\citep{Winn2014}. 

While only $\sim$20\% of \emph{Kepler} planet host stars are multiple transiting planet systems (MTPS), the total number of planets in MTPS accounts for almost half of the \emph{Kepler} planet candidates.~\citet{Latham2011} compared \emph{Kepler} MTPS to single transiting planet systems (STPS). They found a lack of gas giant planets in MTPS, which indicates that the existence of a gas giant planet may disrupt the orbital inclinations or suppress the formation of multiple planets. Furthermore, other studies implied that the distributions of orbital spacings~\citep{Xie2014}, eccentricities~\citep{Xie2015} and obliquities~\citep{Morton2014} are different for STPS and MTPS. In this paper, we investigate one possibility that causes the different orbital architecture between STPS and MTPS, namely, the influence of dynamically-bound companion stars. 

By comparing stellar multiplicity rate for \totaltargets MTPS against stars in the solar neighborhood~\citep{Raghavan2010,Duquennoy1991}, \citet{Wang2014a} found evidence of suppressive planet formation in multiple stellar systems with stellar separations smaller than 20 AU. {{Beyond 20 AU, the stellar multiplicity rate was difficult to measure without high resolution and deep imaging data that provide sensitivity to stellar companions at these separations.}} Therefore, at separations wider than 20 AU, the influence of stellar companions on multi-planet formation was not well understood. {{In this paper, we gather adaptive optics (AO) images for the same MTPS sample in \citet{Wang2014a}. Since AO images for \myaototalno MTPS are already available from the \emph{Kepler} Community Follow-up Observation Program\footnote{https://cfop.ipac.caltech.edu} (CFOP), we obtain new AO images for the remaining \myaototal MTPS at Keck observatory and Palomar observatory. The archival and newly obtained AO images}} reveal dozens of new stellar companions to planet host stars and put valuable constraints on multi-planet formation in multiple stellar systems. 

The paper is organized as follows. We describe the sample selection and AO data acquisition in \S \ref{sec:Sample}, followed by data analyses in \S \ref{sec:Search}. We report the stellar multiplicity rate for MTPS in \S \ref{sec:planet_frequency}. Discussion and summary are given in \S \ref{sec:discussion}. 

\section{Sample Description and AO Data Acquisition}
\label{sec:Sample}

\subsection{Sample Description}
The sample of MTPS remains the same as that in \citet{Wang2014a}. From the NASA Exoplanet Archive\footnote{http://exoplanetarchive.ipac.caltech.edu}, we select \emph{Kepler} Objects of Interest (KOIs) that satisfy the following criteria: (1), disposition of either Candidate or Confirmed; (2), with at least two planet candidates; (3), \emph{Kepler} magnitude ($K_P$) brighter than 13.5. The above selection criteria resulted in \totaltargets MTPS in \citet{Wang2014a}. With the updated Exoplanet Archive, the selection criteria resulted in 208 MTPS. In this paper, we focus {{on}} the \totaltargets MTPS to be consistent with previous work. Their stellar and orbital parameters can be found in Table 2 and Table 3 in \citet{Wang2014a}. 

{{Most MTPS in our sample are true planetary systems based on a statistical analysis by \citet{Lissauer2012}. Subsequent papers on \emph{Kepler} MTPS validated 851 planet candidates in 340 systems~\citep{Rowe2014,Lissauer2014}, 66 MTPS in our sample are included in those validated systems. Furthermore, 25 additional MTPS in our sample are confirmed planetary systems and the remaining 47 MTPS have disposition of planet candidate according to the latest NASA Exoplanet Archive. Therefore, the false positive rate for the MTPS sample studied in this paper should be extremely low. }}

\subsection{AO Data Acquisition}

\subsubsection{Archival AO Data For Follow-up Observations}
\label{sec:archival}

We checked the continually updated CFOP. To avoid repeated AO observations, we only observed KOIs that did not received AO follow-up observations. {{Some of the KOIs without AO data may have speckle imaging~\citep[e.g., ][]{Horch2012,Horch2014} or lucky imaging data~\citep[e.g., ][]{LilloBox2012,LilloBox2014}, but we re-observed these KOIs at Palomar and Keck Observatory because near infrared AO images provide deeper sensitivity and/or higher spatial resolution. For the same reason, we re-observed KOIs that have been observed by the Robo-AO project~\citep{Law2014}. For those KOIs whose AO data from Palomar, MMT, or Keck telescope were available through CFOP, we used the archival AO data. In total, AO data for \myaototalno KOIs were obtained from CFOP and AO data for \myaototal KOIs were obtained by new observations at Palomar and Keck observatory. }}

\subsubsection{AO Imaging with PHARO at Palomar}
\label{sec:pharo}

We observed \myaopalomar KOIs in the sample with the PHARO instrument~\citep{Brandl1997,Hayward2001} at the Palomar 200-inch telescope {{(San Diego County, California, United States)}}. The observations were made between UT July 13rd and 17th in 2014 with seeing varying between 1.0$^{\prime\prime}$ and 2.5$^{\prime\prime}$. PHARO is behind the Palomar-3000 AO system, which provides a on-sky Strehl of 86\% in $K$ band~\citep{Burruss2014}. The pixel scale of PHARO is 25 mas pixel$^{-1}$. With a mosaic 1K $\times$1K detector, the field of view (FOV) is 25$^{\prime\prime}\times$25$^{\prime\prime}$. We normally obtained the first image in $K$ band with a 5-point dither pattern, which had a throw of 2.5$^{\prime\prime}$. AO images in $K$ band provide higher sensitivity to bound companions with late spectral type than $J$ and $H$ band images. {{Furthermore, the AO correction in $K$ band is better and offers a better characterized point spread function (PSF). This is because image quality improves towards longer wavelength for a given wavefront sensing and correcting error~\citep{Davies2012}. A better image with a more stable PSF facilitates companion detection and characterization. }}Exposure time was set such that the peak flux of the KOI is at least 10,000 ADU for each frame, which is within the linear range of the detector. {{If a stellar companion was detected, we observed the KOI in $J$ and $H$ bands right after the $K$ band observation. The color information is useful for estimating the stellar properties of the stellar companion and determining whether the companion is physically bound (see \S \ref{sec:phy_ass}). Nearly simultaneous $J$, $H$, and $K$ band observations help to minimize the influence of any time variability of the target.}} 

\subsubsection{AO Imaging with NIRC2 at Keck II}
\label{sec:nirc2}

We observed \myaokeck KOIs in the sample with the NIRC2 instrument~\citep{Wizinowich2000} at the Keck II telescope {{(Mauna Kea, Hawaii, United States)}}. The observations were made on UT July 18th and August 18th in 2014 with excellent/good seeing between 0.3$^{\prime\prime}$ to 0.8$^{\prime\prime}$. NIRC2 is a near infrared imager designed for the Keck AO system. We selected the narrow camera mode, which has a pixel scale of 10 mas pixel$^{-1}$. The FOV is thus 10$^{\prime\prime}\times$10$^{\prime\prime}$ for a mosaic 1K $\times$1K detector. {{We started the observation in $K$ band for each KOI for the same reason stated in \S \ref{sec:pharo} and followed by $J$ and $H$ band observations if any stellar companions were found.}} The exposure time setting is the same as the PHARO observation: we ensured that the peak flux is at least 10,000 ADU for each frame. We used a 3-point dither pattern with a throw of 2.5$^{\prime\prime}$. We avoided the lower left quadrant in the dither pattern because it has a much higher instrumental noise than the other 3 quadrants on the detector.

\section{Data Analyses}
\label{sec:Search}
\subsection{Contrast Curve and Detections}
\label{sec:contrast_curve}

The raw data were processed using standard techniques to replace bad pixels, subtract dark, flat-field, subtract sky background, align and co-add frames. {{We constructed a bad pixel map using dark frames. Pixels with dark current that deviated more than 5-$\sigma$ from their surrounding pixels were recorded as bad pixels. Their values were replaced with the median flux of surrounding pixels. Dark frames were obtained with the exact same setting as the science frames, e.g., exposure time, co-adds, and read-out mode. After dark subtraction, each science frame was corrected for flat fielding. The dithered science frames provided an estimate of the sky background which was subtracted off from the science frames. The dark-subtracted, flat-fielded, sky-removed science frames were then co-added, resulting in a single frame for subsequent analyses. }}

\begin{figure}[htp]
\begin{center}
\includegraphics[angle=0, width= 1.0\textwidth]{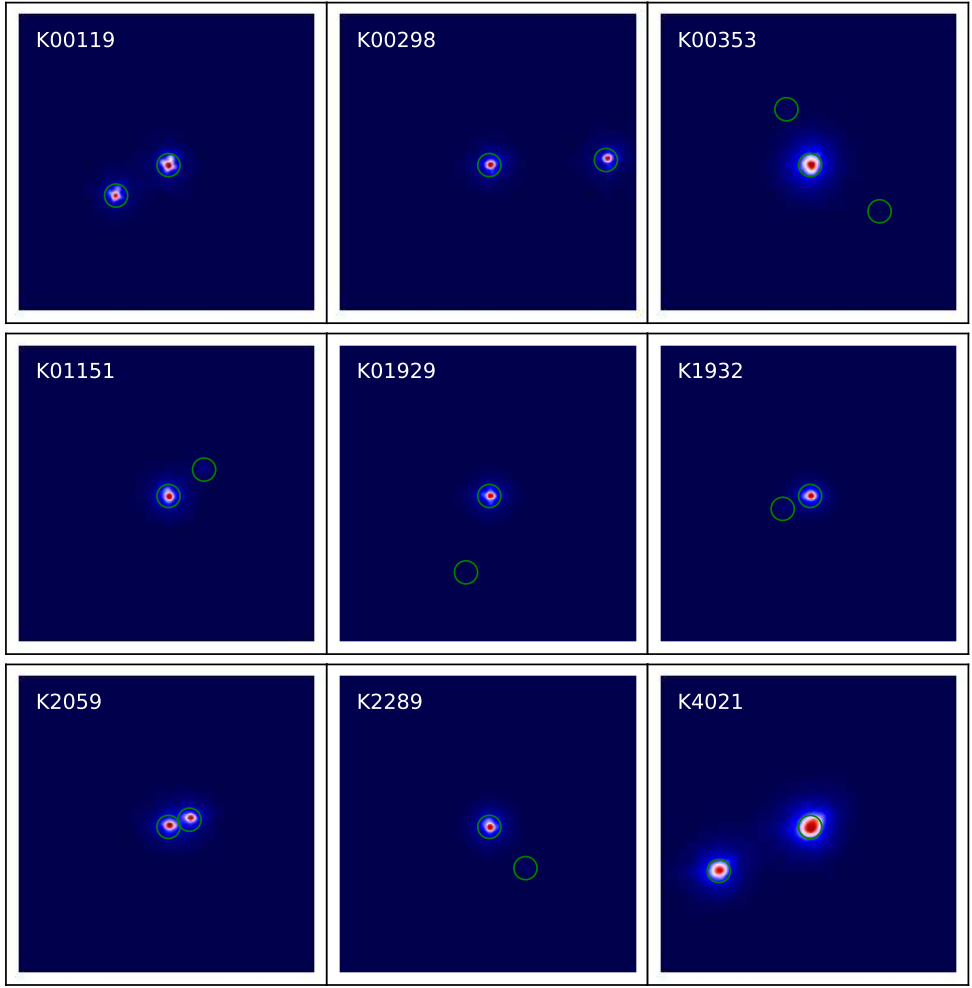} 
\caption{AO images for 9 KOIs with newly detected stellar companions within 2$^{\prime\prime}$. All images cover a 2$^{\prime\prime}$ by 2$^{\prime\prime}$ sky region centering at the primary star. North is up and east is to the left. Linear color scale is chosen such that the central star (red) is normalized to 1 and the background (blue) represents 1/100 of the central star flux. Both central stars and detected stellar companions are marked by green circles. Photometric and astrometric information of detected stellar companions can be found in Table \ref{tab:AO_detections}.
\label{fig:ao_detections}}
\end{center}
\end{figure}

We calculated 5$\sigma$ detection limit as follows. We defined a series of concentric annuli centering on the star. For the concentric annuli, we calculated the median and the standard deviation of flux for pixels within these annuli. We used the value of five times the standard deviation above the median as the 5$\sigma$ detection limit. The detection limits at different angular separations are reported in Table \ref{tab:ao_params}. {{We developed an automatic program to detect stellar companions whose differential magnitudes are brighter than the 5-$\sigma$ detection limit. The program recorded the differential magnitude, position, position angle, detection significance of each detection. All detections were then visually checked to remove confusions such as speckles, background extended sources, and cosmic ray hits.}} In total, \detectstar stellar companions were detected within 5$^{\prime\prime}$ around \detectsys KOIs. Their properties are summarized in Table \ref{tab:AO_detections}. {{Fig. \ref{fig:ao_detections} shows 9 KOIs with newly detected stellar companions within 2$^{\prime\prime}$.}}

\subsection{Physical Association}
\label{sec:phy_ass}

For stellar companions detected by imaging techniques, we need to check whether they are optical doubles/multiples, which will systematically increase the stellar multiplicity rate. To test physical association, \citet{Ngo2015} obtained multiple-epoch AO images and measured common proper motion. In our case, \emph{Kepler} stars are generally further away and common proper motion is more difficult to measure. Given only one epoch of observation, we can use color information of detected stellar companions and assess the probability of their physical association to primary stars~\citep{LilloBox2014,Wang2014b,Wang2015}. {{The color information provides an estimate of the stellar properties, which can then be used to estimate distance for consistency check between the primary and the secondary stars. Any inconsistent distance would be an indication that the primary and the secondary stars are optical doubles. }}For stellar companions with only single-band observations, color information is not available. We can assess the probability with a galactic stellar population simulation. This method is described in detail in~\citet{Wang2015} and the physical association probabilities of each detected stellar companions are given in Table \ref{tab:AO_detections}. 

\subsection{Combining AO Observations with Other Techniques}
\label{sec:aorvda}

\begin{figure}[htp]
\begin{center}
\includegraphics[angle=0, width= 0.9\textwidth]{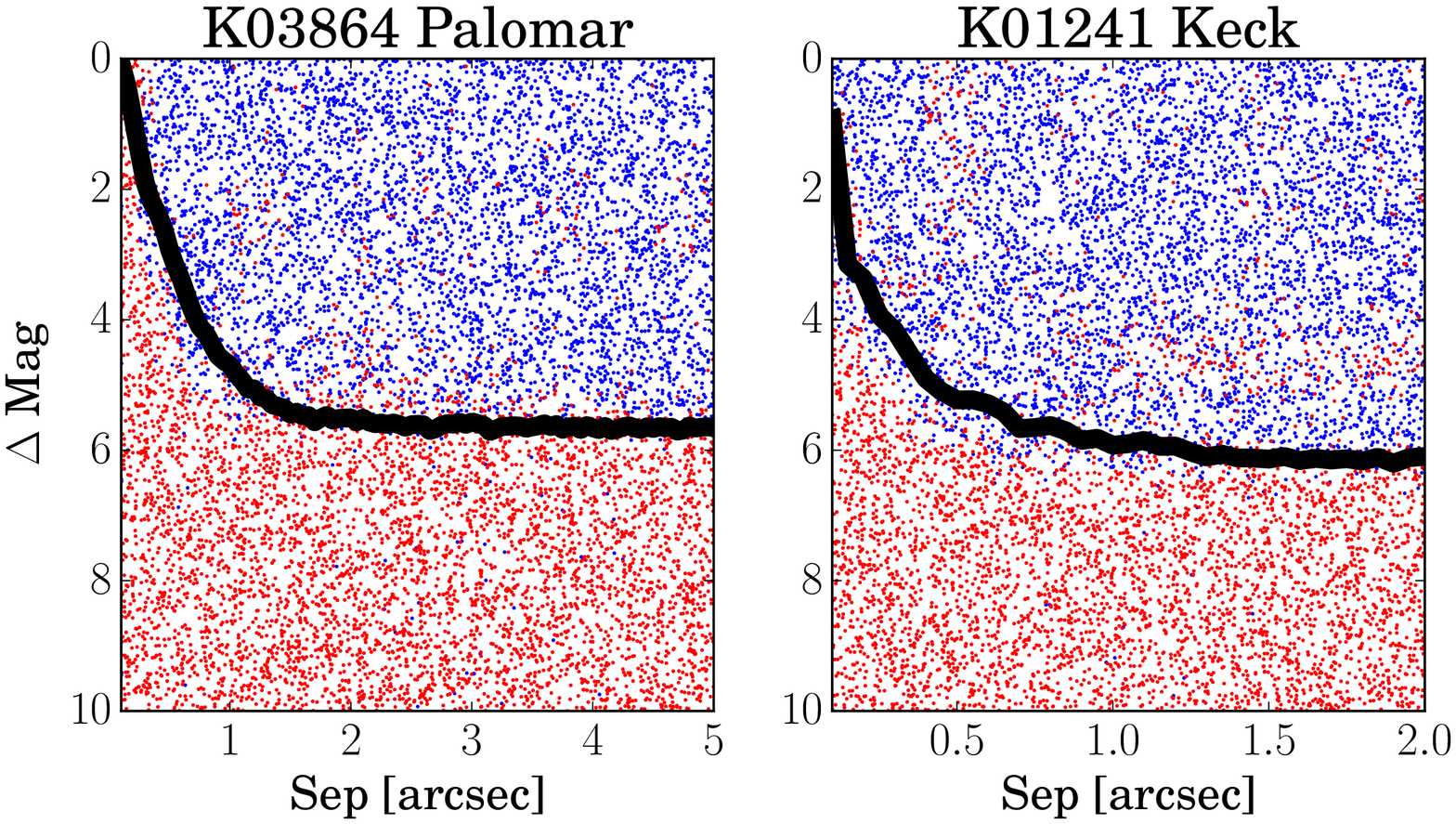} 
\caption{Simulation for AO search completeness in comparison with contrast curve. Left panel shows an example for a Palomar AO image and right panel for a Keck AO image. Blue dots are artificial PSF injections at random separations, differential magnitudes and position angles that are successfully recovered by our detection pipeline. Red dots are injections that are missed. AO contrast curves (\S \ref{sec:contrast_curve}) are plotted as black solid lines which generally trace the border line between blue and red dots. 
\label{fig:fits_injection}}
\end{center}
\end{figure}

Following the method described in~\citet{Wang2015}, we conduct simulations to estimate the search completeness for the AO observations. {{In these simulations, we use the AO contrast curve as a threshold for detection. In practice, however, not all stars above the AO contrast curve are detected by our pipeline, so we run another simulation to test the goodness of using the contrast curve as a threshold. The simulation is identical to other studies~\citep{Gilliland2014,LilloBox2014,Ngo2015} that artificially inject companion stars with the same PSF at random separations, differential magnitudes and position angles. The results are shown in Fig. \ref{fig:fits_injection} for two examples, one for a Palomar AO image and the other one for Keck. For the Palomar AO image, 94.7\% of injected companion stars above the contrast curve are successfully recovered by our detection pipeline and 88.2\% of injections below the contrast curve are missed. For the Keck image, 90.7\% of injections are recovered above the contrast curve and 88.4\% are missed below the contrast curve. The simulation shows that using the contrast curve as a detection threshold is a reasonable assumption. The resulting AO search completenesses are within a few percent for the case of using AO contrast curve as a hard limit  for detection and for the case using the artificial PSF injection result~\citep{Gilliland2014,LilloBox2014,Ngo2015}. The comparable results are due to a relatively smooth distribution of masses and separations of stellar companions, which translates to a smooth distribution on the $\Delta \rm{Mag}$ - angular separation plane as shown in Fig. \ref{fig:fits_injection}. The hard-edge effect of using the AO contrast curve is averaged out and becomes comparable with a more realistic artificial PSF injection simulation.   }}

Since AO imaging technique is not sensitive to stellar companions within or close to the diffraction limit of a telescope, we use other techniques to constrain the presence of stellar companions, i.e., the RV technique and the dynamical analysis~\citep{Wang2014a}. There are \rvstar KOIs in our sample with at least 3 epochs of RV observation. Following the description of \citet{Wang2014b}, we use the Keplerian Fitting Made Easy (KFME) package~\citep{Giguere2012} to analyze the RV data. Among \rvstar KOIs with RV data, only KOI-5 exhibits a RV trend. The stellar companion that can potentially induce the trend is constrained to be beyond 7 AU~\citep{Wang2014b}. More recent RV data suggest that, in addition to two transiting planet candidates, two more distant components exist in KOI-5 system (Howard Isaacson, private communication). One is a sub-stellar companion with a period of $\sim$2700 days and the other one is the AO-imaged stellar companion. Therefore, we consider the closest stellar companion to KOI-5 has a projected separation of 40.12 AU (Table \ref{tab:AO_detections}).

Besides RV and AO observations, we can use dynamical analysis to put additional constraints on potential stellar companions. This dynamical analysis makes use of the co-planarity of MTPS discovered by the \emph{Kepler} mission~\citep{Lissauer2011}. A stellar companion with high mutual inclination to the planetary orbits would have perturbed the orbits and significantly reduced the co-planarity of planetary orbits, and hence the probability of multi-planet transits{{~\citep[see \S 2.6 in][]{Wang2014a}}}. Therefore, the fact that we have observed multiple transiting planet helps to exclude the possibility of a highly-inclined stellar companion. The dynamical analysis is complementary to the RV technique because it is sensitive to stellar companions with large mutual inclinations to the planetary orbits. {{For systems with no stellar companions detected by the AO and/or RV method, an isolation probability can be calculated based on the search completeness of AO and RV observations and the constraints from the dynamical analysis~\citep{Wang2015}. The isolation probability is a measure of how likely a star is isolated from other stellar companions within a certain distance. The isolation probabilities within 2000 AU for KOIs with non-detections of stellar companions are given in Table \ref{tab:ao_params}.  }} 

\section{Stellar Multiplicity Rate For MTPS}
\label{sec:planet_frequency}

\begin{figure}[htp]
\begin{center}
\includegraphics[angle=0, width= 0.9\textwidth]{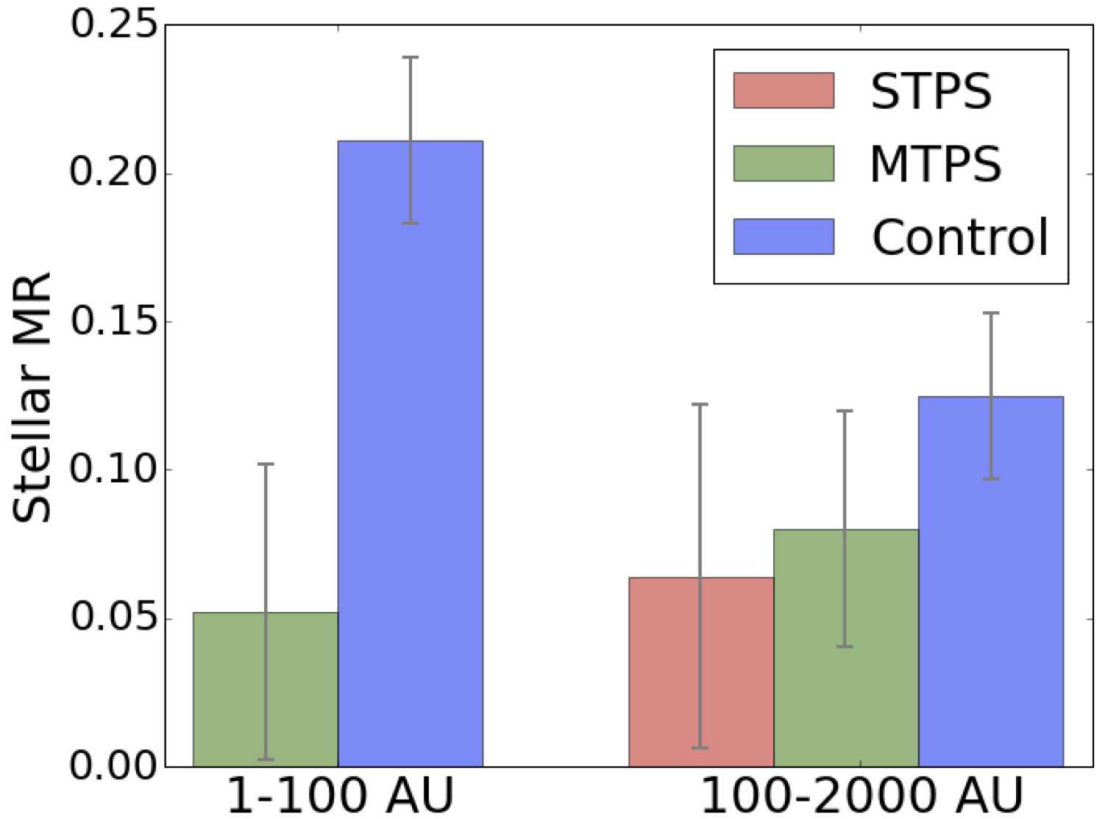} 
\caption{ 
Stellar multiplicity rate for multiple transiting planet systems (MTPS, green), single transiting planet systems (STPS, red), and the field stars in the solar neighborhood, i.e., the control sample in blue. The stellar multiplicity rates for different samples are given in Table \ref{tab:smr}.
\label{fig:mr}}
\end{center}
\end{figure}

Following the same method described in~\citet{Wang2015}, we calculate the stellar multiplicity rate for MTPS as a function of $a$, i.e., companion semi-major axis. We find that for $1\ \rm{AU}<a<2000$ AU, the stellar multiplicity rate for MTPS is $13.3\pm5.7$\%, which is significantly (3.2$\sigma$) lower than $33.6\pm2.8$\% for the control sample, i.e., the field stars in the solar neighborhood~\citep{Raghavan2010}. We choose an upper limit of 2000 AU for comparison because the separation roughly corresponds to the smallest field of view of co-added AO images, which have the best sensitivity for stellar companion search. We further divide the semi-major axis of a stellar companion into two ranges, $1\ \rm{AU}<a<100$ AU and $100\ \rm{AU}<a<2000$ AU. We choose 100 AU because of two reasons. First, the separation is roughly the effective range of the perturbation of coplanarity by a companion star (see discussion of \S \ref{sec:comp_single}). Second, 100 AU is roughly the borderline of RV and AO sensitivity~\citep{Wang2014a,Wang2014b}. Beyond 100 AU, the AO sensitivity is much higher than that for the RV technique. The stellar multiplicity rates for MTPS are $5.2\pm5.0$\% and $8.0\pm4.0$\% for $1\ \rm{AU}<a<100$ AU and $100\ \rm{AU}<a<2000$ AU, respectively. In comparison, the stellar multiplicity rates are $21.1\pm2.8$\% and $12.5\pm2.8$\% for the control sample in these two stellar separation ranges. The stellar multiplicity rate of MTPS for $1\ \rm{AU}<a<100$ AU is lower (2.8$\sigma$) than that for the control sample. For $100\ \rm{AU}<a<2000$ AU, the stellar multiplicity rates are comparable between MTPS and the control sample. Fig. \ref{fig:mr} illustrates the comparison of the stellar multiplicity rates in these two separation ranges. 

\section{Discussion and Summary}
\label{sec:discussion}

\subsection{Interpretation of the Stellar Multiplicity of MTPS}
\label{sec:interpret}

{{The stellar multiplicity rate for MTPS ($5.2\pm5.0$\%) is 2.8$\sigma$ lower than that for stars in the solar neighborhood ($21.1\pm2.8$\%) for $1\ \rm{AU}<a<100$ AU. }}The difference may result from two {{possible}} origins {{that are not mutually exclusive}}. First, MTPS occur less frequently in multiple stellar systems. {{Suppressive planet formation in multiple stellar systems has been noted in previous observational works on both RV and transiting planet samples~\citep[e.g., ][]{Eggenberger2011,Roell2012,Wang2014a} and recently a theoretical work~\citep{Touma2015}. However, other works suggest that the influence of a stellar companion may not be significant~\citep{Gilliland2014,Horch2014} or may be facilitative depending on the stellar separation and planetary mass~\citep{Wang2015,Ngo2015}. 

If suppressive planet formation does not play a role, there may be another origin for the low stellar multiplicity rate: MTPS are less likely to be observed in multiple stellar systems~\citep{Wang2014a}. Coplanarity of MTPS can be affected by an additional stellar component. Thus the likelihood of observing multiple transiting planets is reduced.}} 

If suppressive planet formation plays a major role, then our measurements of stellar multiplicity rates indicate that within 100 AU, MTPS occur less frequently due to the influence of stellar companions. {{For $100\ \rm{AU}<a<2000$ AU, since the stellar multiplicity rates are comparable (0.9$\sigma$ difference) between MTPS ($8.0\pm4.0$\%) and the control sample ($12.5\pm2.8$\%), we conclude that the influence of stellar companions, if any, is too small to be observed.}}

\subsection{Comparison to STPS}
\label{sec:comp_single}

If coplanarity is responsible for the observed low stellar multiplicity rate for MTPS, then we should expect a difference of stellar multiplicity rate between MTPS and STPS. Note that the influence of stellar companions on coplanarity depends on stellar separations. If stellar separations are beyond $\sim$100 AU, their influence on coplanarity is negligible~\citep{Wang2014a,Wang2014b}. Therefore, any difference of stellar multiplicity rate beyond 100 AU is more likely to be due to the origin of planet formation rather than the companions' influence on coplanarity.

In \ref{sec:interpret}, we show that beyond 100 AU the stellar multiplicity rates are comparable between MTPS and the control sample. Here, we compare MTPS to STPS. {{Since these two populations likely have different dynamical history~\citep{Xie2014,Morton2014}, the comparison allows us to study whether the difference is related to stellar multiplicity. }}

From CFOP, we select 89 \emph{Kepler} STPS. The selection criteria are the same as described in \S \ref{sec:Sample} with two exceptions: 1, the number of transiting planet is equal to one; 2, they must have AO images. The stellar properties of these STPS are given in Table \ref{tab:stellar_params}. The sample of these STPS is a subsample of \emph{Kepler} stars with high-resolution imaging observations from CFOP~\citep{Ciardi2015}. Out of these 89 \emph{Kepler} stars, only 6 have RV observations. Since the RV technique is sensitive to close-in stellar companions, obtaining the statistics for stellar companions within 100 AU is difficult. Therefore, we focus on $100\ \rm{AU}<a<2000$ AU. The AO detections are listed in Table \ref{tab:AO_detections_single}. Following the same method in \citet{Wang2015}, we find that the stellar multiplicity rate is  $6.4\pm5.8$\% for STPS for $100\ \rm{AU}<a<2000$ AU,. The value is consistent with that for MTPS, i.e., $8.0\pm4.0$\%. Therefore, we find no evidence that stellar companions between 100 and 2000 AU are responsible for the difference of orbital configuration between MTPS and STPS. However, the difference may be caused by stellar companions within 100 AU, for which we do not have adequate observational constraints. 

\subsection{Comparison to Previous Result}
\label{sec:comp_PaperII}

The same sample of \totaltargets MTPS were studied in~\citet{Wang2014a}. {{They found evidence of suppressive planet formation in tight binary stellar systems with $a<$ 20 AU. }}This finding is consistent with the finding in this paper that the stellar multiplicity rate for MTPS is lower than the control sample within 100 AU at 2.8$\sigma$ level. However, we cannot rule out another possibility that may cause the low stellar multiplicity, i.e., the influence of stellar companions on coplanarity of planetary orbits.

Combining newly obtained AO imaging data with archival RV data, we improve the statistics of stellar companions of planet host stars at large semi-major axes. For example, in \citet{Wang2014a}, stellar multiplicity rate can only be constrained within $\sim$100 AU because of a lack of AO imaging data. In this work, we extend the constraints to 2000 AU. Even within 100 AU, the stellar companion statistics is improved by the AO imaging data. This is because the AO imaging technique complements the RV technique at semi-major axes at which the dynamical signals are difficult to detect. The combination of AO and RV data enables the detection of a deficit of stellar companions to MTPS within 100 AU. 

\citet{Wang2014b} combined RV and AO data for 56 \emph{Kepler} planet host stars. The stellar multiplicity rate for $a<2000$ AU was $43.2\pm5.7$\%, which is a factor of three higher than what we reported in this paper, i.e., $13.3\pm5.7$\%. The discrepancy is due to two reasons. First, we exclude optical doubles whereas \citet{Wang2014b} included both optical doubles and physically associated companions. {{A physical separation of 2000 AU roughly corresponds to 3$^{\prime\prime}$-6$^{\prime\prime}$ angular separation (for the typical distances to these Kepler stars), at which the physical association probability is $\sim$50\%.}} {{Therefore, roughly half of visual companions are expected to be optical doubles around 2000 AU.}} Second, we considered statistics of stellar companions to planet host stars when calculating the incompleteness of companion search~\citep{Wang2015}. In comparison, \citet{Wang2014b} considered statistics of stellar companions for stars in the solar neighborhood. The companion search incompleteness was overestimated in \citet{Wang2014b}  because the stellar multiplicity rate for planet host stars is generally lower than that for stars in the solar neighborhood especially for small semi-major axes. Therefore, the correction factor due to search incompleteness is smaller, resulting a lower stellar multiplicity rate. 

\subsection{Summary and Conclusion}
\label{sec:Summary}

We study the influence of stellar companions on MTPS using a sample of \totaltargets \emph{Kepler} MTPS. We search for stellar companions to these planet host stars with AO images and archival RV data. In total, we detected \detectstar stellar companions {{within 5$^{\prime\prime}$}} around \detectsys multi-planet host stars. The properties of detected stellar companions are summarized in Table \ref{tab:AO_detections}. We also provide detection limits for all stars in our sample in Table \ref{tab:ao_params}. 

We compare the stellar multiplicity rate between MTPS and a control sample, i.e., stars in the solar neighborhood. For semi-major axes $1\ \rm{AU}<a<2000$ AU the stellar multiplicity rate is $13.3\pm5.7$\% for MTPS, which is 3.2$\sigma$ lower than $33.6\pm2.8$\% for the control sample, i.e., the field stars in the solar neighborhood~\citep{Raghavan2010}. The deficit of stellar companions to MTPS can be a result of two origins, a suppressive planet formation and the disruption of coplanarity due to stellar companions. Since the latter may only be effective within 100 AU, we divide the semi-major axes into two ranges, $1\ \rm{AU}<a<100$ AU and $100\ \rm{AU}<a<2000$ AU. The stellar multiplicity rate of MTPS for $1\ \rm{AU}<a<100$ AU is lower (2.8$\sigma$) than that for the control sample. The stellar multiplicity rates are comparable between MTPS and the control sample for $100\ \rm{AU}<a<2000$ AU. 

We also compare the stellar multiplicity rates for MTPS and STPS. No quantitative difference is found between MTPS and STPS for $100\ \rm{AU}<a<2000$ AU. For $1\ \rm{AU}<a<100$ AU, our data are insufficient for comparative study between MTPS and STPS because of a lack of RV data for STPS. Based on these results, we cannot distinguish the two origins that could be responsible for the low stellar multiplicity rate for MTPS for $1\ \rm{AU}<a<100$ AU. Future AO and RV follow-up observations for a larger sample are needed for such a comparative study between MTPS and STPS.

\noindent{\it Acknowledgements} The authors thank the anonymous referee for constructive comments and suggestions that greatly improve the paper. We would like to thank the telescope operators and supporting astronomers at the Palomar Observatory and the Keck Observatory. Some of the data presented herein were obtained at the W.M. Keck Observatory, which is operated as a scientific partnership among the California Institute of Technology, the University of California and the National Aeronautics and Space Administration. The Observatory was made possible by the generous financial support of the W.M. Keck Foundation. The research is made possible by the data from the \emph{Kepler} Community Follow-up Observing Program (CFOP). The authors acknowledge all the CFOP users who uploaded the AO and RV data used in the paper. This research has made use of the NASA Exoplanet Archive, which is operated by the California Institute of Technology, under contract with the National Aeronautics and Space Administration under the Exoplanet Exploration Program. J.W.X. acknowledges support from the National Natural Science Foundation of China (Grant No. 11333002 and 11403012), the Key Development Program of Basic Research of China (973 program, Grant No. 2013CB834900) and the Foundation for the Author of National Excellent Doctoral Dissertation (FANEDD) of PR China.
J.W. acknowledges the travel fund from the Key Laboratory of Modern Astronomy and Astrophysics (Nanjing University).

\bibliography{mybib_JW_DF_PH5}

%
%
%
%

%

\clearpage

\newpage
\input{ao_params.tex} 
\newpage
\input{ao_detections.tex} 
\newpage
\input{stellar_params.tex} 
\newpage
\input{ao_detections_single.tex} 
\newpage
\input{stellar_multiplicity.tex} 
\end{document}

%% file: ao_params.tex

\clearpage

\begin{deluxetable}{lcccccccccccccccc}
\tabletypesize{\tiny}
 \setlength{\tabcolsep}{0.00in} 
\tablewidth{0pt}
\tablecaption{AO Sensitivity \label{tab:ao_params}}
\tablehead{
\multicolumn{7}{c}{\textbf{Kepler}} &
\multicolumn{4}{c}{\textbf{Observation}} &
\multicolumn{6}{c}{\textbf{Limiting Delta Magnitude$^{**}$}} \\
\colhead{\textbf{KIC}} &
\colhead{\textbf{KOI}} &
\colhead{\textbf{Kmag}} &
\colhead{\textbf{$i$}} &
\colhead{\textbf{$J$}} &
\colhead{\textbf{$H$}} &
\colhead{\textbf{$K$}} &
\colhead{\textbf{Comp.}} &
\colhead{\textbf{Iso.}} &
\colhead{\textbf{Instrument}} &
\colhead{\textbf{Flt}} &
\colhead{\textbf{0.1}} &
\colhead{\textbf{0.2}} &
\colhead{\textbf{0.5}} &
\colhead{\textbf{1.0}} &
\colhead{\textbf{2.0}} &
\colhead{\textbf{4.0}} \\
\colhead{\textbf{}} &
\colhead{\textbf{}} &
\colhead{\textbf{[mag]}} &
\colhead{\textbf{[mag]}} &
\colhead{\textbf{[mag]}} &
\colhead{\textbf{[mag]}} &
\colhead{\textbf{[mag]}} &
\colhead{\textbf{$<$ 5$^{\prime\prime}$}} &
\colhead{\textbf{Prob.$^{*}$}} &
\colhead{\textbf{}} &
\colhead{\textbf{}} &
\colhead{\textbf{[$^{\prime\prime}$]}} &
\colhead{\textbf{[$^{\prime\prime}$]}} &
\colhead{\textbf{[$^{\prime\prime}$]}} &
\colhead{\textbf{[$^{\prime\prime}$]}} &
\colhead{\textbf{[$^{\prime\prime}$]}} &
\colhead{\textbf{[$^{\prime\prime}$]}} 
}

\startdata

8554498 & 00005 & 11.665 & 11.485 & 10.542 & 10.257 & 10.213 & yes & \nodata &    NIRC2&       K&     2.0&     4.0&     6.4&     7.4&     7.5&     7.5 \\
8554498 & 00005 & 11.665 & 11.485 & 10.542 & 10.257 & 10.213 & yes & \nodata &    PHARO&       J&     0.1&     1.3&     2.7&     4.8&     6.7&     7.6 \\
6521045 & 00041 & 11.197 & 11.030 & 10.081 & 9.804 & 9.768 & no & 0.96 &    ARIES&       K&     0.1&     1.2&     3.4&     5.9&     7.2&     7.5 \\
6521045 & 00041 & 11.197 & 11.030 & 10.081 & 9.804 & 9.768 & no & 0.96 &    NIRC2&       K&     3.2&     4.6&     5.3&     5.4&     5.4&     5.4 \\
6521045 & 00041 & 11.197 & 11.030 & 10.081 & 9.804 & 9.768 & no & 0.96 &    PHARO&       J&     0.4&     3.6&     4.5&     6.6&     7.6&     7.7 \\
6850504 & 00070 & 12.498 & 12.284 & 11.252 & 10.910 & 10.871 & yes & \nodata &    PHARO&       J&     0.4&     3.0&     4.5&     6.4&     7.3&     7.4 \\
11904151 & 00072 & 10.961 & 10.778 & 9.889 & 9.563 & 9.496 & no & 0.99 &    ARIES&       K&     0.8&     2.3&     5.1&     7.0&     7.6&     7.6 \\
11904151 & 00072 & 10.961 & 10.778 & 9.889 & 9.563 & 9.496 & no & 0.99 &    PHARO&       J&     0.5&     3.3&     4.3&     6.5&     7.9&     8.1 \\
\enddata

\tablecomments{$^{*}$: Isolation probability is the probability of a KOI being isolated within 2000 AU (i.e., has no stellar companion within 2000 AU) given the AO and/or RV data and/or dynamical analysis (see \S \ref{sec:aorvda}). For stars with detected nearby stellar companions, the physical association probability can be found in Table \ref{tab:AO_detections}. $^{**}$: Limiting Delta Magnitudes are 5-$\sigma$ limit. }

\end{deluxetable}

\clearpage

%% file: ao_detections.tex

\begin{deluxetable}{lclcccllcccl}
\tabletypesize{\tiny}
\tablewidth{0pt}
\setlength{\tabcolsep}{0.01in}
\tablecaption{Visual companion detections with AO data for \emph{Kepler} MTPS.\label{tab:AO_detections}}
\tablehead{
\colhead{\textbf{KOI}} &
\colhead{\textbf{Star\#}} &
\colhead{\textbf{Telescope}} &
\colhead{\textbf{Filter}} &
\colhead{\textbf{$\Delta$ Mag$^{*}$}} &
\multicolumn{2}{c}{\textbf{Separation$^{**}$}} &
\multicolumn{2}{c}{\textbf{Distance}$^{***}$} &
\colhead{\textbf{PA}} &
\colhead{\textbf{Association$^{****}$}} &
\colhead{\textbf{ref.$^{\dag}$}} \\
\colhead{\textbf{}} &
\colhead{\textbf{}} &
\colhead{\textbf{}} &
\colhead{\textbf{}} &
\colhead{\textbf{}} &
\colhead{\textbf{}} &
\colhead{\textbf{}} &
\colhead{\textbf{Primary}} &
\colhead{\textbf{Secondary}} &
\colhead{\textbf{}} &
\colhead{\textbf{Probability}} &
\colhead{\textbf{}} \\
\colhead{\textbf{}} &
\colhead{\textbf{}} &
\colhead{\textbf{}} &
\colhead{\textbf{}} &
\colhead{\textbf{(mag)}} &
\colhead{\textbf{(arcsec)}} &
\colhead{\textbf{(AU)}} &
\colhead{\textbf{(pc)}} &
\colhead{\textbf{(pc)}} &
\colhead{\textbf{(deg)}} &
\colhead{\textbf{}} &
\colhead{\textbf{}} \\
}

\startdata

K00005 & 1 & Keck & K & 2.20 & 0.14 & 40.12 & $ 286.6^{ 71.1}_{ -15.8}$ & $^{}_{}$ & 307.4 & $>0.90$ & CFOP\\
\hline
K00070 & 1 & Palomar & J & 4.41 & 3.77 & 1052.60 & $ 279.5^{ 25.3}_{ -23.6}$ & $^{}_{}$ & 51.8 & 0.52 & A12\\
\hline
K00102 & 1 & Palomar & J & 1.12 & 2.84 & 934.31 & $ 329.4^{ 75.0}_{ -30.5}$ & $^{}_{}$ & 222.2 & $>0.90$ & A12\\
\hline
K00108 & 1 & Palomar & J & 5.71 & 2.51 & 891.07 & $ 354.6^{ 45.4}_{ -39.2}$ & $^{}_{}$ & 285.2 & 0.48 & A12 \\
\hline
K00108 & 2 & Palomar & J & 5.60 & 3.23 & 1145.12 & $ 354.6^{ 45.4}_{ -39.2}$ & $^{}_{}$ & 100.8 & 0.30 & A12\\
\hline
K00108 & 3 & Palomar & J & 6.60 & 5.00 & 1773.09 & $ 354.6^{ 45.4}_{ -39.2}$ & $^{}_{}$ & 112.5 & 0.00 & A12\\
\hline
K00115 & 1 & MMT & K & 5.06 & 4.00 & 2168.27 & $ 542.1^{ 140.6}_{ -97.0}$ & $^{}_{}$ & 89.7 & 0.33 & A12\\
\hline
K00119 & 1 & Palomar & J & 0.16 & 1.05 & 327.89 & $ 313.0^{ 106.8}_{ -62.2}$ & $ 380.8^{ 499.6}_{ -154.5}$ & 119.1 & $>0.90$ & this work \\
\enddata

\tablecomments{References: A12 - \citet{Adams2012}; C15 - \citet{Campante2015}. $^{*}$: Typical $\Delta$ Mag uncertainty is 0.1 mag. The uncertainty is estimated from the companion injection simulation described in \S \ref{sec:aorvda}.  $^{**}$: Typical angular separation uncertainty is 0.05$^{\prime\prime}$. The uncertainty is estimated from the companion injection simulation described in \S \ref{sec:aorvda}. $^{***}$: Distance is estimated based on stellar properties of primary stars~\citep{Huber2014} and color information of secondary stars~\citep[see \S 4.1 in][ for more details]{Wang2015}. $^{****}$: Association probability has 10\% uncertainty due to statistical error in simulation. $^{\dag}$: AO images from CFOP are provided by David Ciardi unless otherwise noted. }

\end{deluxetable}

%% file: stellar_params.tex
\clearpage

\begin{deluxetable}{lccccccc}
\tabletypesize{\scriptsize}
 \setlength{\tabcolsep}{0.01in} 
\tablewidth{0pt}
\tablecaption{Stellar Parameters For STPS\label{tab:stellar_params}}
\tablehead{
\colhead{\textbf{KOI}} &
\colhead{\textbf{KIC}} &
\colhead{\textbf{$\alpha$}} &
\colhead{\textbf{$\delta$}} &
\colhead{\textbf{$Kp$}} &
\colhead{\textbf{$T_{\rm eff}$}} &
\colhead{\textbf{$\log g$}} &
\colhead{\textbf{[Fe/H]}} \\
\colhead{\textbf{}} &
\colhead{\textbf{}} &
\colhead{\textbf{(h:m:s)}} &
\colhead{\textbf{(d:m:s)}} &
\colhead{\textbf{(mag)}}	&
\colhead{\textbf{(K)}}	&
\colhead{\textbf{(cgs)}}	&
\colhead{\textbf{(dex)}}	
}

\startdata

00042&   8866102&    18:52:36.17&     45:08:23.4& 9.36& 6325& 4.26& 0.01\\
00069&   3544595&    19:25:40.39&    38:40:20.49& 9.93& 5669& 4.47&-0.18\\
00084&   2571238&    19:21:40.99&    37:51:06.48&11.90& 5543& 4.57&-0.14\\
00087&  10593626&     19:16:52.2&    47:53:04.06&11.66& 5642& 4.44&-0.27\\
00092&   7941200&    18:53:29.96&    43:47:17.59&11.67& 5952& 4.49&-0.04\\
00103&   2444412&       19:26:44&    37:45:05.73&12.59& 5653& 4.55&-0.06\\
00118&   3531558&    19:09:27.07&    38:38:58.56&12.38& 5747& 4.18& 0.03\\
00122&   8349582&    18:57:55.79&    44:23:52.95&12.35& 5699& 4.17& 0.30\\
\enddata


\end{deluxetable}

\clearpage

%% file: ao_detections_single.tex

\begin{deluxetable}{lclcccllcccl}
\tabletypesize{\tiny}
\tablewidth{0pt}
\setlength{\tabcolsep}{0.01in}
\tablecaption{Visual companion detections with AO data for \emph{Kepler} STPS.\label{tab:AO_detections_single}}
\tablehead{
\colhead{\textbf{KOI}} &
\colhead{\textbf{Star\#}} &
\colhead{\textbf{Telescope}} &
\colhead{\textbf{Filter}} &
\colhead{\textbf{$\Delta$ Mag$^{*}$}} &
\multicolumn{2}{c}{\textbf{Separation$^{**}$}} &
\multicolumn{2}{c}{\textbf{Distance}$^{***}$} &
\colhead{\textbf{PA}} &
\colhead{\textbf{Association$^{****}$}} &
\colhead{\textbf{ref.$^{\dag}$}} \\
\colhead{\textbf{}} &
\colhead{\textbf{}} &
\colhead{\textbf{}} &
\colhead{\textbf{}} &
\colhead{\textbf{}} &
\colhead{\textbf{}} &
\colhead{\textbf{}} &
\colhead{\textbf{Primary}} &
\colhead{\textbf{Secondary}} &
\colhead{\textbf{}} &
\colhead{\textbf{Probability}} &
\colhead{\textbf{}} \\
\colhead{\textbf{}} &
\colhead{\textbf{}} &
\colhead{\textbf{}} &
\colhead{\textbf{}} &
\colhead{\textbf{(mag)}} &
\colhead{\textbf{(arcsec)}} &
\colhead{\textbf{(AU)}} &
\colhead{\textbf{(pc)}} &
\colhead{\textbf{(pc)}} &
\colhead{\textbf{(deg)}} &
\colhead{\textbf{}} &
\colhead{\textbf{}} \\
}

\startdata

K00118 & 1 & Palomar & J & 3.94 & 1.24 & 583.76 & $ 470.3^{ 18.8}_{ -24.4}$ & $ 1152.1^{ 878.0}_{ -605.8}$ & 214.3 & $>0.90$ & CFOP\\
K00118 & 1 & Palomar & K & 3.65 & 1.23 & 578.94 & $ 470.3^{ 18.8}_{ -24.4}$ & $ 1152.1^{ 878.0}_{ -605.8}$ & 214.6 & $>0.90$ & CFOP\\
\hline
K00268 & 1 & MMT & J & 3.03 & 1.57 & 372.64 & $ 238.1^{ 32.6}_{ -7.1}$ & $ 305.4^{ 116.2}_{ -274.7}$ & 179.7 & $>0.90$ & CFOP\\
K00268 & 1 & MMT & K & 2.52 & 1.65 & 392.07 & $ 238.1^{ 32.6}_{ -7.1}$ & $ 305.4^{ 116.2}_{ -274.7}$ & 174.8 & $>0.90$ & CFOP\\
K00268 & 1 & Palomar & K & 2.47 & 1.75 & 415.62 & $ 238.1^{ 32.6}_{ -7.1}$ & $ 305.4^{ 116.2}_{ -274.7}$ & 267.3 & $>0.90$ & CFOP\\
\hline
K00268 & 2 & MMT & J & 4.37 & 2.34 & 556.29 & $ 238.1^{ 32.6}_{ -7.1}$ & $ 305.4^{ 116.2}_{ -274.7}$ & 128.1 & $>0.90$ & CFOP (Dupree)\\
K00268 & 2 & MMT & K & 3.87 & 2.33 & 554.58 & $ 238.1^{ 32.6}_{ -7.1}$ & $ 305.4^{ 116.2}_{ -274.7}$ & 132.0 & $>0.90$ & CFOP (Dupree)\\
K00268 & 2 & Palomar & K & 3.72 & 2.49 & 593.65 & $ 238.1^{ 32.6}_{ -7.1}$ & $ 305.4^{ 116.2}_{ -274.7}$ & 309.9 & $>0.90$ & CFOP\\
\enddata

\tablecomments{$^{*}$: Typical $\Delta$ Mag uncertainty is 0.1 mag. The uncertainty is estimated from the companion injection simulation described in \S \ref{sec:aorvda}.  $^{**}$: Typical angular separation uncertainty is 0.05$^{\prime\prime}$. The uncertainty is estimated from the companion injection simulation described in \S \ref{sec:aorvda}. $^{***}$: Distance is estimated based on stellar properties of primary stars~\citep{Huber2014} and color information of secondary stars~\citep[see \S 4.1 in][ for more details]{Wang2015}. $^{****}$: Association probability has 10\% uncertainty due to statistical error in simulation. $^{\dag}$: AO images from CFOP are provided by David Ciardi unless otherwise noted. }

\end{deluxetable}

%% file: stellar_multiplicity.tex

\begin{deluxetable}{ccccccc}
\tablewidth{0pt}
\tablecaption{Stellar multiplicity rate within a certain stellar separation for MTPS, STPS, and field stars in the solar neighborhood (i.e., the control sample) .\label{tab:smr}}
\tablehead{
\colhead{\textbf{$a$}} &
\multicolumn{2}{c}{\textbf{MTPS}} &
\multicolumn{2}{c}{\textbf{STPS}} &
\multicolumn{2}{c}{\textbf{Control Sample}} \\
\colhead{\textbf{[AU]}} &
\colhead{\textbf{MR}} &
\colhead{\textbf{$\delta$MR}} &
\colhead{\textbf{MR}} &
\colhead{\textbf{$\delta$MR}} &
\colhead{\textbf{MR}} &
\colhead{\textbf{$\delta$MR}} \\
}

\startdata

$1<a<100$ & 0.052	& 0.050	&     \nodata	&     \nodata	&     0.211 & 0.028\\
$100<a<2000$ & 0.080	& 0.040	&     0.064 &     0.058	&     0.125 & 0.028\\

\enddata


\end{deluxetable}